\documentclass[apj]{emulateapj}

\usepackage{color}

\renewcommand\email\texttt
\newcommand\bmath[1]{\ensuremath{\mbox{\boldmath $#1$}}}

\newcommand\kms{\mbox{km s$^{-1}$}}
\newcommand\kmss{\mbox{km$^2$ s$^{-2}$}}
\newcommand\masyr{\mbox{mas yr$^{-1}$}}
\newcommand\tilt{\alpha_{ij}}
\newcommand\tiltrt{\alpha_{r\theta}}
\newcommand\tiltrp{\alpha_{r\phi}}
\newcommand\tiltpt{\alpha_{\phi\theta}}

\begin{document} 

\slugcomment{\sc submitted to \it the Astrophysical Journal}

\shorttitle{Tilt of the Halo Velocity Ellipsoid} \title{The Tilt of
  the Halo Velocity Ellipsoid and the Shape of the Milky Way Halo}

\shortauthors{Smith et al.}
\author{Martin C. Smith,\altaffilmark{1}
N. Wyn Evans,\altaffilmark{1}
and Jin H. An\altaffilmark{2,3}}

\altaffiltext{1}{Institute of Astronomy, University of Cambridge,
Madingley Road, Cambridge CB3 0HA, UK; \email{msmith@ast.cam.ac.uk,nwe@ast.cam.ac.uk}}
\altaffiltext{2}{Dark Cosmology Centre, Niels Bohr Institute, 
University of Copenhagen, Juliane Maries Vej 30,
DK-2100 Copenhagen \O, Denmark \email{jin@dark-cosmology.dk}}
\altaffiltext{3}{Niels Bohr International Academy, Niels Bohr Institute,
University of Copenhagen, Blegdamsvej 17, DK-2100 Copenhagen \O, Denmark}

\begin{abstract}
  A sample of $\sim$1,800 halo subdwarf stars with radial velocities
  and proper motions is assembled from Bramich et al.'s (2007)
  light-motion catalog. This is based on the repeated multi-band Sloan
  Digital Sky Survey photometric measurements in Stripe 82. Our sample
  of halo subdwarfs is extracted via a reduced proper motion diagram
  and distances are obtained using photometric parallaxes, thus giving
  full phase space information. The tilt of the velocity ellipsoid
  with respect to the spherical polar coordinate system is computed
  and found to be consistent with zero for two of the three tilt
  angles, and very small for the third.  We prove that if the inner
  halo is in a steady-state and the triaxial velocity ellipsoid is
  everywhere aligned in spherical polar coordinates, then the
  potential must be spherically symmetric. The detectable, but very
  mild, misalignment with spherical polars is consistent with the
  perturbative effects of the Galactic disk on a spherical dark
  halo. Banana orbits are generated at the 1:1 resonance (in
  horizontal and vertical frequency) by the disk. They populate
  Galactic potentials at the typical radii of our subdwarf sample,
  along with the much more dominant short-axis tubes. However, on
  geometric grounds alone, the tilt cannot vanish for the banana
  orbits and this leads to a slight, but detectable, misalignment. We
  argue that the tilt of the stellar halo velocity ellipsoid therefore
  provides a hitherto largely neglected but important line of argument
  that the Milky Way's dark halo, which dominates the potential, must
  be nearly spherical.
\end{abstract}

\section{Introduction}

The kinematics of any stellar population are often most conveniently
described by its velocity dispersion tensor
\begin{equation}
\sigma^2_{ij} \equiv
\left\langle
(v_i - \langle v_i \rangle)(v_j - \langle v_j \rangle)
\right\rangle
\label{eq:cov}
\end{equation}
where the subscript indices denote one of the orthogonal coordinate
directions, and the angled brackets represent averaging over the phase
space distribution function \citep[see e.g.,][]{Bi08}. The dispersion
tensor is a symmetric second-rank tensor and so may always be
diagonalized. The principal axes of the tensor then form a velocity
ellipsoid, which need not be aligned with the coordinate
directions. However, as already realized by \citet{Ed15}, if the
gravity field is time-independent, then the alignment of the velocity ellipsoid
is a powerful global probe of the gravitational potential.

The triaxiality of the local halo velocity ellipsoid is
well-established \citep[e.g.,][]{Wo78,Ch00,Go03,Ke07,Sm09}. However,
despite its importance, the alignment of the velocity ellipsoid of
halo stars has received very little attention. Here, we construct an
unprecedentedly large sample of $\sim$1,800 halo subdwarf stars with
known distances, radial velocities and proper motions in Sloan Digital
Sky Survey (SDSS) Stripe 82. We find that the velocity dispersion
tensor is anisotropic, with an alignment very close to that of the
spherical polar coordinate axes.

If the velocity ellipsoid is exactly aligned radially everywhere, then
the potential of the Milky Way is spherical in shape. The very mild
misalignment that we detect is consistent with the influence of the
Galactic disk on an underlying spherical Galactic halo potential.

\section{The Subdwarf Catalog}

\subsection{Sample Construction}

We construct our sample of subdwarfs using data from the sixth SDSS
data release \citep{Ad08}, in particular utilizing the light-motion
catalog of \citet{Br08}. This is built from the multi-epoch,
multi-band ($u,g,r,i,z$) photometry available for one of the SDSS
equatorial stripes (Stripe 82) and covers $\sim 250~\mbox{deg$^2$}$
in the right ascension range $20\fh7 < \alpha < 3\fh3$ and in the
declination range $|\delta| < 1\fdg26$. A full description of our
subdwarf sample is given elsewhere \citep{Sm09}. Here, we give a brief
outline of the selection procedure.

The reduced proper motion is defined as
\begin{equation}
H_r = r + 5\, \log_{10} \left( \frac{\mu}{\masyr} \right) - 10,
\end{equation}
where $\mu$ is the proper motion and $r$ is the apparent magnitude in
the $r$ band, corrected for extinction using the maps of
\citet{Sc98}. The reduced proper motion is useful because it is 
independent of distance, namely
\begin{equation}
\label{eq:h_mock}
H_r = M_r + 5\, \log_{10} \left( \frac{v_{\tan}}{4.74~\kms} \right).
\end{equation}
where $M_r$ is the absolute magnitude in the $r$ band and $v_{\tan}$ is the
tangential component of the velocity with respect to the line-of-sight
between the Sun and the star.  The reduced proper motion diagram is a
plot of $H_r$ versus color $g-i$, in which the populations of disk
dwarf stars, white dwarfs and the halo subdwarfs are all separated
(see e.g., \citealt{Vi07}, who have already constructed a reduced proper
motion diagram for Stripe 82 to isolate ultracool and halo white
dwarfs).

Here, we are interested in selecting a clean sample of halo subdwarfs,
and so we apply two pre-selection cuts in order to reduce the
contamination.\footnote{Possible contaminants such as white dwarfs,
  disk dwarfs or background giants can be considered to be negligible;
  \citet{Sm09} conclude that the level of contamination for this
  sample is $\la 1\%$.} First, we only use stars that pass a quality
cut such that uncertainties in the proper motion are smaller than
$4~\masyr$. Secondly, the $r$ magnitude of the star should be brighter
than $r = 19.5$. This gives us a sample size of 372,811.  The latter
cut allows us to remove interlopers which may be at large distances
and hence have small $H_r$ despite having $\mu \sim 0~\masyr$. Neither
cut introduces any kinematic bias. Note that we avoid cutting on $\mu$
directly in order to simplify our calculation of the detection
efficiency.

\begin{figure}
\begin{center}
\includegraphics[width=\hsize]{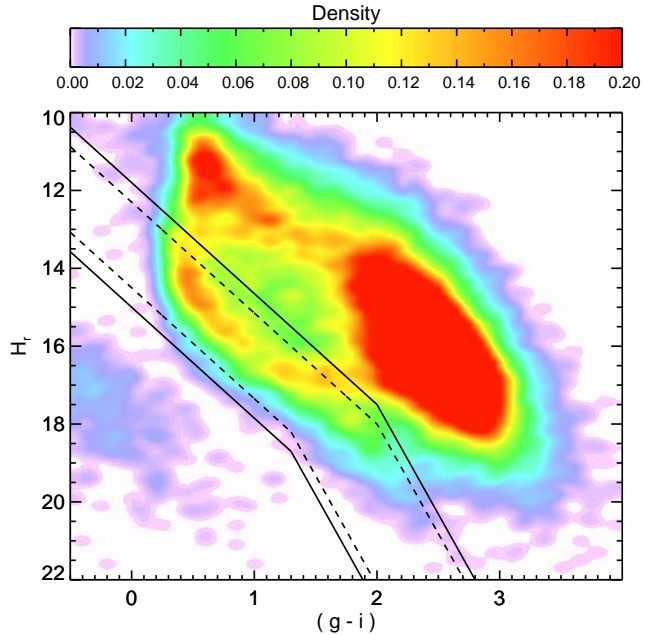} 
\caption{\label{fig:rpm} Reduced proper motion diagram of Stripe 82,
  where the color-scale corresponds to number density (scaled so that
  the peak is unity). The solid lines show the location of our
  subdwarf boundary, while the dashed lines show the region adopted to
  reduce contamination. To improve the clarity of the figure, we have
  incorporated a cut on the proper motion ($\mu > 30~\masyr$); the
  sample used throughout the paper has no such cut on $\mu$. Note that
  the color-scale saturates at 20\% of the peak density.}
\end{center}
\end{figure}

In Figure~\ref{fig:rpm}, we show the reduced proper motion
diagram. Owing to larger tangential velocities, the halo subdwarfs are
clearly differentiated from the slower moving disk dwarfs. We estimate
the location of the subdwarf boundary as
\begin{eqnarray}
  H_r < 2.85 (g-i) + 11.8 &\ \mbox{for }& (g-i) \le 2 \nonumber \\
  H_r < 5.63 (g-i) + 6.24 &\ \mbox{for }& (g-i) > 2 \nonumber \\
  H_r > 2.85 (g-i) + 15.0 &\ \mbox{for }& (g-i) \le 1.3 \nonumber \\
  H_r > 5.63 (g-i) + 11.386 &\ \mbox{for }& (g-i) > 1.3,
\end{eqnarray}
and further clean the sample by rejecting all objects, which are
within $\Delta H_r = 0.5$ of the boundary. These cuts result in a
sample of 30,760 stars. We then cross-match with the output from the
SDSS SEGUE spectral parameter pipeline \citep{Le08}. This provides us
with the radial velocities for members of the subdwarf sample (with
median error of $4.6~\kms$), although it significantly reduces the
sample size to 2,210 stars. The SEGUE spectroscopic target selection
is a complicated function. Suffice it to say that, despite its
complexity, it is believed to be free of any significant kinematic
biases.

In order to recover the full six-dimensional phase-space information,
we must determine the distance to each subdwarf. \citet{Iv08} have
already constructed a photometric parallax relation from SDSS
observations of clusters, obtaining an intrinsic scatter of
$\sim0.2~\mbox{mag}$ \citep{Iv08,Ju08,Se08}. Their parallax relation
is also a function of [Fe/H], which we obtain from the spectral parameter
pipeline (with median error of 0.14 dex). Overall, the errors
obtained from the \citet{Iv08} relation are impressively small, aided
by the high precision photometry from \citet{Br08} -- for our sample
the median relative distance error is 10.1\%.

In conjunction with the good accuracy on the \citet{Br08} proper
motions (less than $4~\masyr$), we are able to construct an
unprecedentedly large sample of halo subdwarfs with accurate positions
and kinematics. The median error on each of our velocity components is
$\sim 30$-$40~\kms$. Restricting ourselves to subdwarfs with
heliocentric distances less than 5 kpc gives a final sample of 1,782.
These stars lie at Galactocentric cylindrical polar radii between 7
and 10 kpc, and at depths of 4.5 kpc or less below the Galactic plane.

\subsection{Kinematic Bias Quantification}

A drawback to our sample is that it is not kinematically unbiased, as
the cut in the reduced proper motion diagram is implicitly a function
of the kinematics. Therefore, we have to model and understand the
effect of this if we are to investigate the distributions of
velocities in our sample.  Notice that the kinematic bias comes solely
from our cut on the reduced proper motion $H_r$ -- which actually
selects stars via their tangential velocity rather than their proper
motion. This makes the task of quantifying the bias significantly
easier since we do not need to make any assumptions about the
underlying distance distribution (i.e., luminosity function).

We calculate our detection efficiency as follows. For each subdwarf in
our final sample, we take the sky coordinates and create a mock sample
of 50,000 fake stars. Then, for each mock star, we select $M_r$ and
$(g-i)$ at random from our observed distributions. Note that for each
realization, $both$ the magnitude and color are assigned from one
star, i.e., we do not assign an absolute magnitude from one star and a
color from another. We then select kinematics for each mock star using
the Galactocentric halo velocity distributions from \citet{Ke07} but
with no net rotation~\citep{Al06}. To calculate $v_{\tan}$ from the
Galactocentric velocities requires us to assign a distance to each
mock star, which we do at random from the observed distribution. This
means that the efficiency does have a dependence on the distance
distribution. However, this dependence is very mild since equation
(\ref{eq:h_mock}) is a function of the tangential velocity rather than
the proper motion.  The efficiency is then given by the fraction of
mock stars which pass our reduced proper motion cut.

In order to check whether our results are dependent on the assumed
halo velocity distribution, we repeat the calculations using the
values from \citet{Ke07}, but now assuming a rotational velocity of
the halo of $\sim 20~\kms$ (in the direction of disk rotation). We
find that uncertainties in the efficiency make little difference to
the final determination of the tilts.


\section{The Alignment of the Halo Velocity Ellipsoid}

\subsection{Method}

With the efficiencies in hand, we can now calculate the misalignment
of the velocity ellipsoid of the SDSS subdwarfs. To do this, we
transform our subdwarf velocities into Galactocentric spherical polar
coordinates: $v_r$ is the radial velocity with respect to the center
of the Galaxy, $v_\theta$ is the zenithal component measured from the
North Galactic Pole, and $v_\phi$ is the azimuthal component measured
such that the Galactic rotation has negative $v_\phi$.

The misalignment from the spherical polar coordinate surfaces can then
be described by the correlation coefficients and the tilt angles using
the following formula
\begin{equation}
{\rm Corr}[v_i,v_j] = 
\frac{ \sigma_{ij}^2 }{ (\sigma_{ii}^2\sigma_{jj}^2)^{1/2} }
\end{equation}
and
\begin{equation}
\tan(2\tilt) = 
\frac{ 2\sigma_{ij}^2 }{ \sigma_{ii}^2 - \sigma_{jj}^2 }.
\label{eq:tiltdef}
\end{equation}
Here the tilt angle corresponds to the angle between the $i$-axis and
the major axis of the ellipse formed by projecting the three
dimensional velocity ellipsoid onto the $ij$-plane.  (see e.g., Binney
\& Merrifield 1998, or Appendix A of this paper). We use the tilt
angles to specify the orientation of the velocity ellipsoid, instead
of alternatives such as the Euler angles because the former are a
natural extension of the familiar two dimensional case and much easier
to visualize than other options. See also recent examples of using the
tilt angles in similar context by Dehnen \& Binney (1998) and Siebert
et al. (2008).

The measured sample covariance is due to both the true underlying
covariance and the correlated measurement uncertainties, i.e.,
\begin{equation}
{\rm Cov}_{\rm m}[v_i, v_j]= \sigma_{ij}^2 + {\rm Cov}[\delta v_i, \delta v_j],
\end{equation}
where ${\rm Cov}_{\rm m}[v_i, v_j]$ is the covariance as measured from
the sample and ${\rm Cov}[\delta v_i, \delta v_j]$ is the covariance of
the error distributions. To account for the detection efficiency, we
also calculate the sample covariances weighted by the
inverse efficiency, e.g.,
\begin{equation}
{\rm Cov}_{\rm m}[v_i,v_j] = \frac{1}{W} 
\sum_{k=1}^N w_k\,
\bigl(v_{i,k} - \langle v_{i,k}\rangle\bigr)\,
\bigl(v_{j,k} - \langle v_{j,k} \rangle\bigr),
\label{eq:scov}
\end{equation}
where
\begin{displaymath}
W
=1-\sum_{k=1}^N w_k^2
\end{displaymath}
and the summation is over the number of subdwarfs in our final
sample. The weights $w_k$ are proportional to reciprocal of the
efficiency $\propto\epsilon_k^{-1}$ (normalized to unity such that
$\sum_{k=1}^N w_k=1$).

As indicated by equation (\ref{eq:scov}),
when the covariances are calculated we subtract the mean velocities,
i.e., our results are not dependent on the assumed reference
frame. However, as can be seen from \citet{Sm09} our sample displays
no significant net motion.

\begin{figure*}
\begin{center}
\includegraphics[width=\hsize]{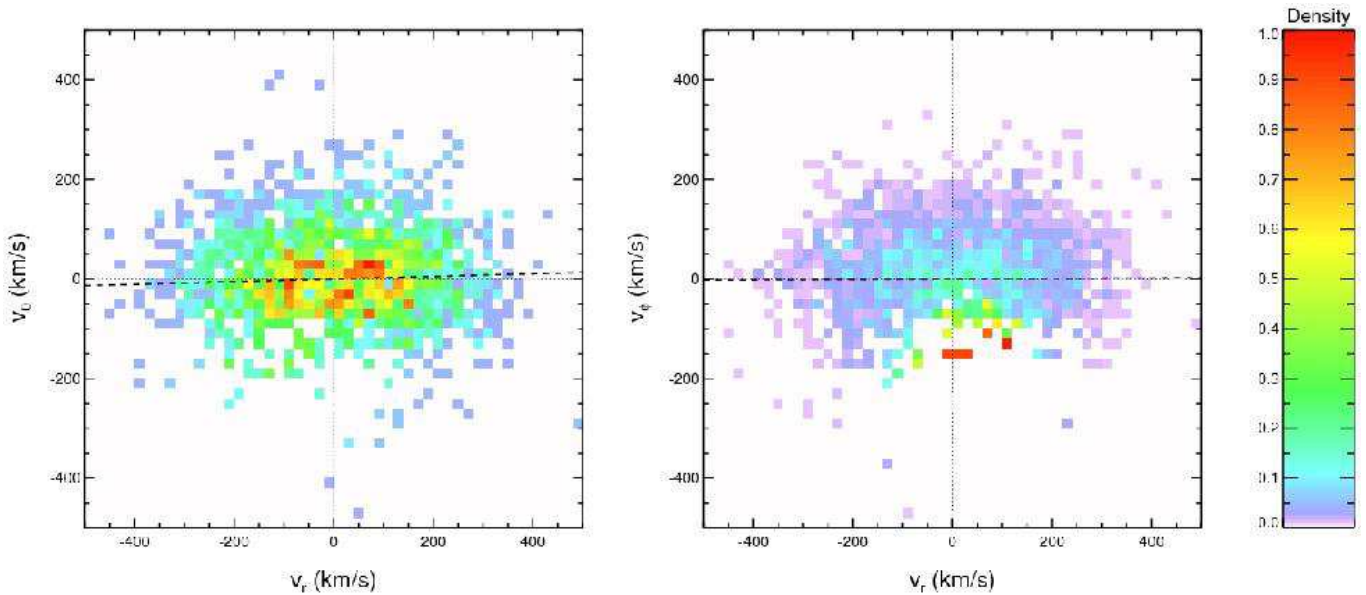} 
\caption{\label{fig:tilt} The efficiency corrected velocity
  distributions in the $(v_r,v_\theta)$ and $(v_r,v_\phi)$ planes for
  the sample of 1,532 subdwarfs with $1~\mbox{kpc} < |z| < 4~\mbox{kpc}$.
  The dashed lines show the orientation of the tilts. The apparent
  non-Gaussianity in the $(v_r,v_\phi)$ distribution is due to the
  variation of the efficiency correction across this plane; for
  certain regions of velocity space (particularly for $v_\phi \approx
  -200~\kms$), our efficiency is low and hence the few stars that fall
  in this range are corrected by a large amount. This results in the
  narrow peaks in density in this figure, although when averaged over
  a larger area the density is consistent with a Gaussian
  distribution.}
\end{center}
\end{figure*}
\begin{figure}
\begin{center}
\includegraphics[width=\hsize,angle=90]{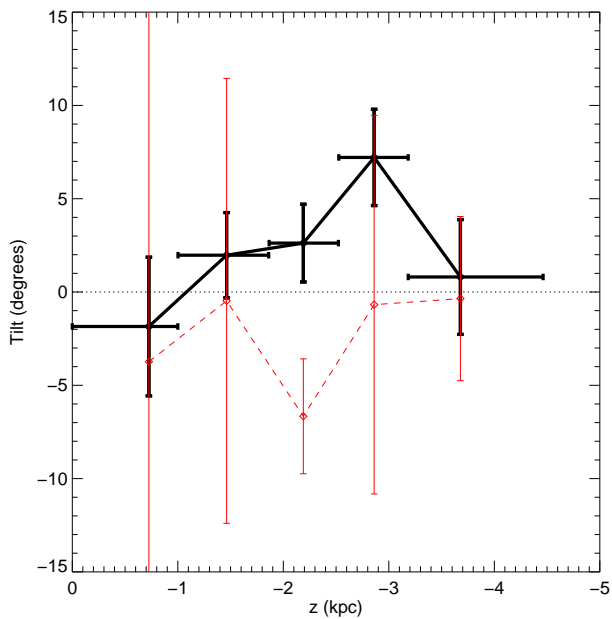} 
\caption{\label{fig:tiltvary} The variation of the tilt angles
  $\tiltrt$ (black) and $\tiltrp$ (red) as a function of height $z$
  from the Galactic plane. The vertical error bars are 1-$\sigma$,
  whilst the horizontal error bars give the bin width. Although there
  are one or two locations where the deviation from the mean is
  greater than 1-$\sigma$, there are no obvious trends discernible.}
\end{center}
\end{figure}

\subsection{The Tilt Angles and Correlations}

Let us first look at the tilt angles in the sample ignoring any stars
with height $|z| < 1~\mbox{kpc}$, so as to minimize (although not
eliminate) any effect due to the Galactic disk potential. This gives
us a sample of 1,532 subdwarfs. We find that the correlations and
tilts are
\begin{eqnarray}
{\rm Corr} [v_r, v_\theta] &= 0.078\!\pm\!0.029,\quad&
\tiltrt= 3\fdg4\!\pm\!1\fdg3
\nonumber \\
{\rm Corr} [v_r, v_\phi] &= -0.028\!\pm\! 0.039,\quad&
\tiltrp = -2\fdg2\!\pm\!3\fdg3
\\
{\rm Corr} [v_\phi, v_\theta] &= -0.087\!\pm\! 0.047,\quad&
\tiltpt = -37\fdg4\!\pm\!20\fdg4
\nonumber
\end{eqnarray}
where the errors are obtained using the bootstrap technique. We find
no evidence of any clear tilt in the $\tiltrp$ and $\tiltpt$
terms. However, the tilt angle $\tiltrt$ is measured to be non-zero at
about 3-$\sigma$ level, though it is still very small. The good
alignment of the velocity ellipsoid in spherical polars is also
apparent from the velocity distributions in the ($v_r, v_\theta$) and
($v_r,v_\phi$) planes illustrated in Figure~\ref{fig:tilt}, in which
the dashed lines show the orientation of the tilts.

Note that the tilt angle $\tiltpt$ is not well constrained since the
dispersions in the two components are similar and the covariance is
small. Hence, the calculation of $\tiltpt$ involves the division of
one small number by another small number, giving a large
error. However, this term is not so important for the purpose of
constraining the shape of the Galactic potential (see the next section).

The area of sky covered by the Stripe 82 catalogue is 249 deg$^2$. It
is reasonable to ask whether the velocity dispersion ellipsoid locally
could have a different behaviour than that averaged over a larger
region, especially as this effect has been observed recently in
numerical simulations~\citep{Ze09}. To study this, we investigate the
variation in tilt as a function of height from the plane for $|z| \le
4.5~\mbox{kpc}$ in Figure~\ref{fig:tiltvary}.  Neither of the tilt
angles $\tiltrt$ and $\tiltrp$ show any significant trends. Although
there are one or two isolated points at which the tilt angles lie at
$\sim 1$-$\sigma$ from the mean, they are localized in longitude,
indicative probably of kinematic substructure and streams which are
expected in the stellar halo in hierarchical assembly models. The
nature of the substructure in the SDSS subdwarfs is addressed in
detail elsewhere \citep{Sm09}. We have also carried out this test by
splitting our sample into three according to right ascension, and find
that the values of the tilt angles are consistent with those derived
for the full sample.

\section{Discussion}

\subsection{Spherical Alignment}

The alignment of the velocity ellipsoid of halo stars in spherical
polar coordinate has substantial implications for the overall
potential of the Galaxy. This constraint does not come from the Jeans
equations, which merely require that the momentum flux balances the
gravitational forces (see \citet{Ev09} or \citet{An09} for recent
applications).  Rather, the constraint comes from the deeper requirement
that a phase space distribution function must exist.  The theorem has
been known for some time, although its widespread applicability has
been obscured by the fact that \citet{Ed15} and \citet{Ch39}
introduced unnecessary assumptions in its proof, as realized first by
\citet{LB62}.

Let us first note that there are a number of trivial ways that permit
the velocity dispersion tensor to be aligned in spherical polar
coordinates. The simplest is to ask for the distribution function to
depend on energy $E$ alone, in which case the velocity dispersion
tensor is everywhere isotropic. Or, we could insist that the
distribution function is given as $f= f(E,|\bmath{L}|)$ in a spherical
potential, or $f= f(E,L_z)$ in an axisymmetric potential. Here,
$\bmath{L}$ is the angular momentum, whilst $L_z$ is the component of
$\bmath{L}$ that is parallel to the symmetry axis.  All these options are not
available to us because of the well-known and long established
triaxiality of the velocity dispersion of halo stars with
$\sigma_{rr}^2 > \sigma_{\phi\phi}^2 > \sigma_{\theta\theta}^2$ -- see
for example recent determinations by \citet{Ke07} or \citet{Sm09}.

To generate the observed triaxial anisotropy of the velocity
dispersion tensor, the phase space distribution function must depend
on at least three independent integrals of motion $I_i$, one of which
may be chosen to be the same as the Hamiltonian, $I_1 =
E(v_r^2,v_\theta^2, v_\phi^2; r,\theta, \phi)$.  Here, it is assumed
that the reference frame is chosen such that there is no net bulk
streaming motion, that is, $\langle v_r\rangle=\langle
v_\theta\rangle=\langle v_\phi\rangle=0$.  Therefore,
$\sigma_{rr}^2=\langle v_r^2\rangle$, $\sigma_{r\theta}^2=\langle
v_rv_\theta\rangle$ and so on.  Accordingly, if the cross-terms
$\langle v_rv_\theta\rangle$, and $\langle v_rv_\phi\rangle$ vanish
everywhere, then any additional integrals of motion that isolate the
distribution function must be even functions of $v_r$. This is
equivalent to the statement that the $v_r$-dependence of the integrals
is only through the square of the radial velocity component $v_r^2$,
and independent of the sign of $v_r$.  Then, the second isolating
integral $I_2$ can be recast as a globally defined function
independent of $v_r^2$ using the energy integral, that is $I_2 =
I_2(E;v_\theta, v_\phi; r, \theta, \phi)$. However, since $I_2$ is an
integral of motion, its Poisson bracket with the Hamiltonian must
vanish, from which it follows that $I_2$ also must be independent of
$r$ as well \citep{LB62}. Hence, the integral can always be cast in
the form $I_2= I_2(E; v_\theta, v_\phi; \theta, \phi)$ and so is
completely independent of both $r$ and (its conjugate momentum) $v_r$.

This implies that the radial coordinate in the Hamilton-Jacobi
equation must separate and so the potential has to be of the form
\begin{equation}
\psi(r,\theta,\phi) = \psi_0(r) + {\Psi(\theta,\phi) \over r^2},
\label{eq:crux}
\end{equation}
where $\psi_0$ and $\Psi$ are arbitrary functions of the indicated
arguments.  In fact, by exactly the same line of reasoning applied to
the third isolating integral $I_3$, if $\langle v_\theta v_\phi\rangle$ also
vanishes, then the potential has to have the form
\begin{equation}
\psi(r,\theta,\phi) = \psi_0(r) + \frac{\xi(\theta)}{r^2} +
\frac{\zeta(\phi)}{r^2 \sin^2\!\theta},
\end{equation}
although this is a stronger result than we will need here.

If the potential has the form~(\ref{eq:crux}), then Poisson's equation
implies that the total density of stars and dark matter is
\begin{equation}
\rho(r,\theta,\phi) = \rho_0(r) + \frac{\Omega(\theta,\phi)}{4\pi Gr^4},
\end{equation}
where
\begin{displaymath}
\rho_0=\frac1{4\pi Gr^2}\frac d{dr}\left(r^2\frac{d\psi_0}{dr}\right)
\end{displaymath}
\begin{displaymath}
\Omega=\frac1{\sin\theta}\frac{\partial}{\partial\theta}
\left(\sin\theta\,\frac{\partial\Psi}{\partial\theta}\right)
+\frac1{\sin^2\!\theta}\frac{\partial^2\Psi}{\partial\phi^2}.
\end{displaymath}
That is to say, the dipole potential is necessarily associated with an
astrophysically unrealistic density cusp diverging as $r^{-4}$ unless
$\Omega=0$.  Consequently, we have $\Psi=0$.\footnote{Mathematically,
  $\Psi$ may be any combination of spherical harmonics
  $Y^l_m(\theta,\phi)$ with $l=1$.  However, there must not be any
  physical source for the pure-dipole gravitational field, and so
  $\Psi$ may be set to zero by appropriate choice of the coordinate
  origin.}  This leaves us with the theorem that:
\begin{quotation} {\it\noindent If a steady state stellar population
    has a non-degenerate (i.e. triaxial) velocity dispersion tensor
    whose eigenvectors are everywhere aligned in spherical polar
    coordinates, then the underlying gravitational potential must be
    spherically symmetric.}
\end{quotation}
This theorem is implicit in \citet{LB62}, whereas restricted versions
were known to \citet{Ed15} and \citet{Ch39}.  In fact, from the
preceding proof, the crux of the result lies in the existence of the
second integral that forces the separation of the radial part of the
Hamilton-Jacobi equation, and the presence of the third integral is of
a secondary importance. Hence, the theorem can be relaxed somewhat:
\begin{quotation} {\it\noindent If the potential is non-singular, it
    is a sufficient condition for a spherical symmetry that one of the
    non-degenerate eigenvectors of the velocity dispersion tensor is
    aligned radially everywhere.}
\end{quotation}
This follows, as spherical symmetry is guaranteed for non-singular
potentials once the radial part of the Hamilton-Jacobi equation can be
separated off.

Of course, spherical alignment of the velocity dispersion tensor does
not imply that the stellar density itself is spherically symmetric. In
fact, many investigators have already found that
$\sigma_{\theta\theta}^2 \neq \sigma_{\phi\phi}^2$ for samples of halo
stars~\citep[e.g.,][]{Go03,Ke07,Sm09}, which implies that the stellar
halo has a density distribution that is flattened or triaxial even
though the gravity field is nearly spherical. This is the case, for
example, in the models presented by \citet{Wh85}, \cite{Ar90}, and
\citet{Ev97}, in which the phase space distribution function depends
on energy and the all three components of angular momentum, $f = f(E,
\bmath L)$.

\begin{figure*}
\begin{center}
\includegraphics[width=0.25\hsize,angle=90]{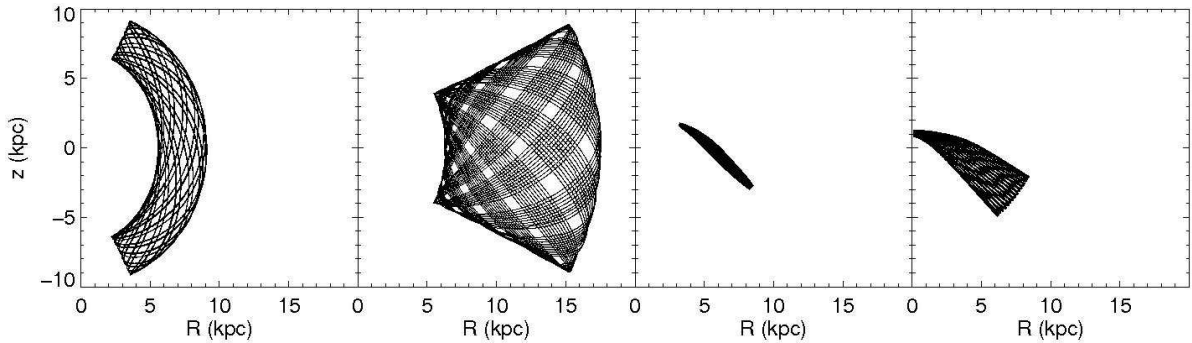} 
\caption{\label{fig:orbits} Cross-sections in the meridional plane of
  typical orbits of the SDSS subdwarfs; leftmost two panels showing
  thin and thick tube orbits, rightmost two panels showing banana or
  saucer orbits.}
\end{center}
\end{figure*}

\subsection{Nearly Spherical Alignment}

The tilt angles $\tiltrp$ and $\tiltpt$ are consistent with zero, but
our results give a very small, but non-zero measurement of the tilt
angle $\tiltrt$ (at the 3-$\sigma$ level).  In an exactly spherical
potential, the $\langle v_rv_\theta\rangle$ cross-term and hence
$\tiltrt$ must vanish. However, even if the dark halo is spherical,
the Galactic potential is not spherical due to the influence of the
bulge and the disk. Of course, at the distances probed by our sample
of SDSS subdwarfs, it is the halo that dominates the gravitational
potential, whilst the disk gives the main perturbation.

The main consequence of the disk is to convert the planar rosette
orbits of a spherical potential into the short-axis tube orbits of a
mildly oblate potential.  However, in typical axisymmetric Galactic
potentials, the key 1:1 resonance (in cylindrical coordinates $R$ and
$z$) also occurs at Galactocentric radii between 2 and 10 kpc, causing
the axial orbits to become unstable to out-of-plane perturbations, and
siring the family of banana or saucer orbits~\citep[see e.g.,][
Section 3.7.3]{Pf84,Mi89,Sc93,Ev94,Bi08}. It is to these two orbital
families -- the short-axis tubes and the bananas -- that our SDSS
subdwarfs will belong.

We can confirm this by computing orbits in the Milky Way potential of
\citet{Fe06}, which comprises a spherical isothermal halo and a
Miyamoto-Nagai disk.  Our initial conditions were chosen as follows:
since the potential is axisymmetric, we start our orbits off from a
location $x = 8$ kpc, $y=0$ kpc and with $z$ distributed uniformly
from $-1$ to $-4~\mbox{kpc}$ (which is a reasonable approximation of
our observed distribution). Our initial velocities were chosen
according to the trivariate ellipsoidal Gaussian $\sigma_r =
142~\kms$, $\sigma_\phi = 81~\kms$, and $\sigma_\theta = 77~\kms$,
i.e., according to the values derived for our observed sample of halo
subdwarfs by \citet{Sm09}.  This is of course not a {\it bona fide}
distribution function, but just a convenient sampling function to scan
phase space. The orbits were calculated using a fourth order
Runge-Kutta method with a time-step of 0.1 Myr, which results in a
fractional energy change of better than $10^{-6}$ over a typical
orbital period of $\sim 200$ Myr.  The orbits so produced are indeed
predominantly short-axis tubes ($\sim 80\%$), but with a reasonable
mixture of banana orbits ($\sim 20\%$). Some examples are illustrated
in the panels of Figure~\ref{fig:orbits}.

{\it It is immediately apparent on geometrical grounds that the banana
orbits can only give positive contributions to $\langle v_r v_\theta
\rangle$.} By contrast, the short-axis tubes can yield both positive
and negative contributions. Hence, the effect of the Galactic disk is
to create a family of orbits for which the tilt term $\sigma_{r\theta}^2$
cannot exactly vanish.

In fact, it is easy to bolster this geometric argument with a
quantitative calculation. Using our sampling function, we generated 96
initial conditions and integrated the orbits for 1 Tyr, recording the
velocities each time a test particle returned to its initial position
(defined in practice to within a spherical volume of 300 pc radius).  The
enormous timescale is to maximize the number of times a particle
returns to its initial position, so that $\sigma_{r\theta}^2$
for the orbit can be calculated as accurately as possible.
Then, we average the resulting values of $\sigma_{r\theta}^2$
for every crossing and for all of our orbits. This gives
$\langle v_rv_\theta\rangle = 690~\kmss$, which is very
comparable to our observed covariance of $893 \pm 335~\kmss$.

However, if we divide our orbits into banana and short-axis tubes
(banana orbits are defined as those which are asymmetric about $z=0$
or the Galactic plane) and recalculate the average
$\sigma_{r\theta}^2$, then we find that the 19 banana orbits have
$\langle v_rv_\theta\rangle = 1740~\kmss$, while the remaining 77
short-axis tube orbits have $\langle v_rv_\theta\rangle = -140~\kmss$.
The banana orbits indeed contribute almost all of the effect. As a
check, if we repeat this procedure in a spherical potential (i.e.,
without the Galactic disk), we indeed find that $\langle
v_rv_\theta\rangle$ is consistent with zero, as it should be.

A very mild misalignment in the velocity ellipsoid of the SDSS
subdwarfs with respect to spherical polar coordinates is therefore
naturally explained by the effects of the Galactic disk.

\section{Conclusions}

The Milky Way's dark halo dominates the gravity field, and different
datasets have yielded very different values for its flattening.  For
example, analyses of the variation of the thickness of the Galaxy's
gas layer with radius point to an oblate halo with axis ratio $q \sim
0.7$ \citep{Ol01}. This is consistent with the typical values of halo
flattening measured for some external galaxies, using the dynamics of
polar rings or the shapes of the isophotes of the X-ray emission for
early-type galaxies.  On the other hand, \citet{Fe06} reached the very
different conclusion that the dark halo of our own Galaxy must be very
close to spherical, based on dynamical modeling of the bifurcation in
the Sagittarius stream. Finally, \citet{He04} used the kinematic data
of stars in the leading arm of the Sagittarius to argue that the dark
halo is prolate.

Here, we have drawn attention to another -- somewhat neglected --
probe of the shape of the dark halo, namely the tilt of the velocity
ellipsoid of halo stars. In a pure spherical halo, the velocity
ellipsoid of any tracer population -- regardless of whether its
density distribution is flattened or triaxial -- must be aligned in
the spherical polar coordinate system. By contrast, the behavior of
the velocity ellipsoid in flattened halos is much more varied. For
example, \cite{Le85a,Le85b} constructed a number of halo models with
ellipticities between 0.3 and 0.7 using the Schwarzschild method. They
show numerous plots of the behavior of the velocity ellipsoid, from
which it is apparent that it deviates, often very strongly, from
spherical polar alignment.

In this paper, the tilt of the velocity ellipsoid of the Milky Way
halo stars has been measured using an unprecedentedly large sample of
$\sim$1,800 halo subdwarfs extracted from Bramich et al.'s (2008)
light-motion catalog, itself based on the repeated Sloan Digital Sky
Survey observations of Stripe 82. We find that the velocity ellipsoid
of the halo subdwarfs is very closely aligned with the spherical polar
coordinate system. In particular, two of the tilt angles are
consistent with zero, whereas the final tilt angle -- corresponding to
the $\langle v_r v_\theta \rangle$ term -- is very small. We have
shown that this effect is consistent with the influence of the
Galactic disk, which must cause a mild asphericity in the total
potential even if the dark halo itself is spherical. This asphericity
gives rise to a family of orbits, the banana or saucer orbits, which
are typically represented in samples of stars at the locations of the
SDSS subdwarfs. On geometric grounds alone, $\langle v_r v_\theta
\rangle$ does not exactly vanish for banana orbits, and so their small
admixture causes a very mild misalignment.

Our argument here is open to the criticism that we have not shown that
the velocity dispersion tensor is aligned everywhere in spherical
polar coordinates. We have merely shown that the alignment is very
close to spherical polars for halo subdwarfs at heliocentric distances
of $\la 5~\mbox{kpc}$ along the $\sim 250~\mbox{deg$^2$}$ covered by
SDSS Stripe 82. Nonetheless, this is still a striking and unexpected
result over a swathe of Galactic locations that provides a new line of
attack on the awkward question of the shape of the Milky Way's dark
halo.  It would be very interesting to extend our results to samples
of halo stars in different directions. This is particular the case for
high latitude samples of halo stars, for which any effects due to the
Galactic disk are negligible, and so for which alignment with the
spherical polar coordinate system should be perfect.

\acknowledgments The authors wish to thank V.~Belokurov, P.~Hewett,
M.~Juri\'c and an anonymous referee for advice and guidance. The Dark
Cosmology Centre is funded by the Danish National Research
Foundation. MCS acknowledges support from the STFC-funded ``Galaxy
Formation and Evolution'' program at the Institute of Astronomy,
University of Cambridge.

Funding for the SDSS and SDSS-II has been provided by the Alfred
P. Sloan Foundation, the Participating Institutions, the National
Science Foundation, the U.S. Department of Energy, the National
Aeronautics and Space Administration, the Japanese Monbukagakusho, the
Max Planck Society, and the Higher Education Funding Council for
England. The SDSS Web Site is http://www.sdss.org/.

The SDSS is managed by the Astrophysical Research Consortium for the
Participating Institutions. The Participating Institutions are the
American Museum of Natural History, Astrophysical Institute Potsdam,
University of Basel, Cambridge University, Case Western Reserve
University, University of Chicago, Drexel University, Fermilab, the
Institute for Advanced Study, the Japan Participation Group, Johns
Hopkins University, the Joint Institute for Nuclear Astrophysics, the
Kavli Institute for Particle Astrophysics and Cosmology, the Korean
Scientist Group, the Chinese Academy of Sciences (LAMOST), Los Alamos
National Laboratory, the Max-Planck-Institute for Astronomy (MPIA),
the Max-Planck-Institute for Astrophysics (MPA), New Mexico State
University, Ohio State University, University of Pittsburgh,
University of Portsmouth, Princeton University, the United States
Naval Observatory, and the University of Washington.

\appendix

\section{The Tilt Angles and the Velocity Ellipsoid}

After the mean motion has been subtracted, the velocity dispersion is
\begin{equation}
\sigma_{ij}^2=\int\!d^3\bmath v\,v_iv_jf.
\end{equation}
If we consider a coordinate transformation $v_{i'}=\Lambda_{i'j}v_j$
(henceforth the summation convention for repeated indices is used) in
velocity space given by a matrix in SO(3), it is straightforward to
verify that the velocity dispersion transforms as a component of a
tensor, i.e.,
\begin{equation}
\sigma_{i'j'}^2=\int\!d^3\bmath v\,v_{i'}v_{j'}f
=\int\!d^3\bmath v\,(\Lambda_{i'k}v_k)(\Lambda_{j'l}v_l)f
=\Lambda_{i'k}\Lambda_{j'l}\sigma_{kl}^2
\end{equation}
where the Jacobian relating the coordinate transformation is unity.

Note from the definition that the velocity dispersion tensor is
symmetric and its three eigenvalues are all positive definite. Hence,
it is also invertible. Let $(\varsigma_{ij}^2)=(\sigma_{ij}^2)^{-1}$
be the matrix inverse of the velocity dispersion tensor. The velocity
ellipsoid is then defined to be a quadric surface in velocity space
\begin{equation}
\varsigma_{ij}^2v_iv_j=1.
\end{equation}
Next, from the rule involving the inversion of the product of matrices
and the fact $\Lambda_{i'j}$ is an orthogonal matrix (i.e.,
$\Lambda^{-1}_{ij'}=\Lambda_{j'i}$), it is clear that
$\varsigma_{ij}^2$ also transforms as a tensor, that is to say,
$\varsigma_{i'j'}^2=\Lambda_{i'k}\Lambda_{j'l}\varsigma_{kl}^2$. Then,
\[
\varsigma_{ij}^2v_iv_j
=\varsigma_{ij}^2\Lambda^{-1}_{ik'}v_{k'}\Lambda^{-1}_{jl'}v_{l'}
=\Lambda_{k'i}\Lambda_{l'j}\varsigma_{ij}^2v_{k'}v_{l'}
=\varsigma_{k'l'}^2v_{k'}v_{l'},
\]
and thus the equation of the velocity ellipsoid is form-invariant
under orthogonal coordinate transformations. If the primed coordinate
axes are chosen to be aligned to the principal axes of the velocity
ellipsoid, the equation of the ellipsoid in the primed coordinate is
in the canonical form. 
It then follows that $\varsigma_{i'j'}^2$ and subsequently
$\sigma_{i'j'}^2$ are diagonalized in the same primed coordinate
system. In other words, the coordinate transformation to the system
with the basis set given by the principal axes of the velocity
ellipsoid is identical to the one diagonalizing the velocity
dispersion tensor.

Next, we consider the projection of the velocity ellipsoid. Here, we
suppose that the projection is onto the $v_xv_y$-plane, although the
other two follow basically the same procedure. Let us think of a line
with fixed $(v_x,v_y)$ but with varying $v_z$. If $(v_x,v_y)$ falls
within the projection of the ellipsoid, the line intersects the
ellipsoid in two points. On the other hand, if the point $(v_x,v_y)$
is outside the projection in the $v_xv_y$-plane, the line does not
cross the ellipsoid. Following similar logic, we find that the point
$(v_x,v_y)$ is on the projection of the ellipsoid in the
$v_xv_y$-plane if the line running perpendicular to the plane through
the given point is a tangent to the ellipsoid. Now, if the equation of
the ellipsoid is considered as a quadratic equation for $z$ given
fixed $(v_x,v_y)$, the preceding argument indicates that setting its
discriminant to be zero defines the equation of the projection in the
$v_xv_y$-plane.  Therefore, we find the equation of the projection of
the velocity ellipsoid onto the $v_xv_y$-plane;
\begin{equation}
C_{yy}v_x^2-2C_{xy}v_xv_y+C_{xx}v_y^2=\varsigma_{zz}^2,
\end{equation}
where $C_{xx}$, $C_{xy}$, and $C_{yy}$ are the matrix cofactors of
$\varsigma_{xx}^2$, $\varsigma_{xy}^2$, and $\varsigma_{yy}^2$,
respectively. This traces an ellipse in $v_xv_y$-plane.

The projected velocity dispersion tensor corresponding to this ellipse
may be defined analogously to the 3-d case. After some algebra, we
find that
\begin{equation}
\left\lgroup\begin{array}{cc}
C_{yy}/\varsigma_{zz}^2&-C_{xy}/\varsigma_{zz}^2\\
-C_{xy}/\varsigma_{zz}^2&C_{xx}/\varsigma_{zz}^2
\end{array}\right\rgroup^{-1}
=\frac1D\left\lgroup\begin{array}{cc}
C_{xx}&C_{xy}\\C_{xy}&C_{yy}
\end{array}\right\rgroup
=\left\lgroup\begin{array}{cc}
\sigma_{xx}^2&\sigma_{xy}^2\\\sigma_{xy}^2&\sigma_{yy}^2
\end{array}\right\rgroup
\end{equation}
where $D=|\varsigma_{ij}^2|$ is the matrix determinant of
$(\varsigma_{ij}^2)$. In other words, the projection of the velocity
ellipsoid onto the $v_xv_y$-plane is the same as the 2-d velocity
`ellipsoid' calculated with the $2\times2$ $(v_x,v_y)$-submatrix of
the original $3\times3$ velocity dispersion matrix. Consequently, the
tilt angle defined as in equation (\ref{eq:tiltdef}) is the same as
the angle between the principal axis of the ellipse that is the
projection of the velocity ellipsoid onto the $ij$-plane and the
coordinate axis. We note however that this is not the same as the
angle between the coordinate axis and the projection of the principal
axis of the velocity ellipsoid onto the same plane -- that is to say,
the principal axis of the velocity ellipsoid is not necessarily
projected into the principal axis of the projected ellipse.

\end{document}